\documentclass[a4paper,twocolumn]{article}
\usepackage[latin1]{inputenc}
\usepackage{float}
\usepackage{amsmath,amssymb}
\usepackage{theorem}
\usepackage{graphicx}

\begin{document}

\title{\bf Intrinsic angular momentum in general relativity}

\author{Osvaldo M. Moreschi\thanks{Member of CONICET.}
\thanks{ Email: moreschi@fis.uncor.edu}~
\\
\small FaMAF, Universidad Nacional de C\'{o}rdoba\\ 
\small Ciudad Universitaria, 
(5000) C\'{o}rdoba, Argentina.
}
\date{November 26, 2002}
\maketitle
%\tableofcontents

\begin{abstract}
There are several definitions of the notion of angular momentum
in general relativity. However none of them can be said to capture
the physical notion of intrinsic angular momentum of the sources
in the presence of gravitational radiation.
We present a definition which is appropriate for the description
of intrinsic angular momentum in radiative spacetimes.
This notion is required in calculations involving
radiation of angular momentum, as for example is expected in
binary coalescence of black holes.

\end{abstract}

%\pacs{
PACS numbers: 04.20.-q, 04.20.Cv, 04.20.Ha
%}

\section{Introduction}
There is a lot of effort invested in the development 
and construction of large interferometric gravitational wave detectors
such as LIGO, VIRGO, GEO, etc. These observatories are expected
to measure the gravitational waves emitted in relativistic 
astrophysical systems, for example in the coalescence of two
compact objects. At the late stages of such systems one can consider the 
situation of having two black holes or neutron stars in nearly circular
orbits which are shrinking due to the loss, by the emission of gravitational
waves, of energy and angular momentum. 

In order to estimate how important is the radiation mechanism of angular
momentum in these kind of processes, let us consider briefly the situation
of two Newtonian point masses, with mass $m_0$ each, in circular motion
at a distance $r$. Then one can calculate the ratio between 
total angular momentum $J$ and total mass square $M=2m_0$, obtaining
$\frac{J}{M^2}=\sqrt{\frac{r}{16\, r_S}}$; where $r_S$ is the
Schwarzschild radius of masses $m_0$, and we are using geometric
units for which the speed of light and the gravitational constant
have the unit value. 
This means that for orbits for which $r>16\,r_S$, the angular momentum
exceeds the Kerr limit to have a final black hole, as opposed to 
a naked singularity. In other words, some how, one has to account for
the radiation of angular momentum before the final collapse of 
the system, if one is to expect a black hole at the end of the
coalescence. It is probably interesting to mention that according
to Newtonian dynamics, the magnitude of the 
velocity of these particles when they
are in the  $r=16\,r_S$ orbit is $25\%$ of the speed of light;
which indicates that a relativistic description of these systems is required.

Since the angular momentum loss is
crucial for the coalescence mechanism, it is important to have an accurate
description of this process. This leads to the first question which is:
what is the definition of angular momentum in a relativistic system?

There are several difficulties associated to this question. 
To begin with, the spacetime appropriate for the description of a relativistic
isolated system of compact objects is curved and asymptotically flat.
Therefore, since the asymptotic symmetry group is the infinite
dimensional BMS group\cite{Bondi62,Sachs62}, 
the notion of angular momentum in general relativity
will have to deal with the so called problem of supertranslations.
Roughly speaking, among the generators of the BMS group, one can distinguish
a set of 6 rotations, and an infinite set of supertranslations. A normal
4-dimensional subgroup of the BMS group exists which allows for the 
unambiguous definition of total momentum; the so called Bondi momentum.

To give perspective to this problem, let us recall that
in special relativity,
the angular momentum $J^{\tt a  b}$, 
the intrinsic angular momentum $S^{\tt a b}$ and
the linear momentum $P^{\tt a}$  are related by
\begin{equation}
  \label{eq:j}
  J^{\tt a  b}=S^{\tt a b}+R^{\tt a} \, P^{\tt b} - P^{\tt a}\, R^{\tt b} ,
\end{equation}
where ${\tt a,b}$ are numeric spacetime indices and $R^{\tt a}$
represents the translational freedom. 
A rest frame is one in which the momentum $P$ has only timelike
components.
Given a rest reference
frame in Minkowski spacetime one needs to use the spacelike translation
freedom appearing in expression (\ref{eq:j}) in order to single out
the center of mass reference frame. In the center of mass frame
one has $ J^{\tt a  b}=S^{\tt a b}$; that is the total momentum 
coincide with the intrinsic angular momentum.

In general relativity, instead of having just a momentum, we have an
infinite component supermomentum vector (associated with the infinite
generators of supertranslations). Therefore the analog of equation
(\ref{eq:j}) necessarily will forces us to deal with the notion
of supermomentum and its corresponding notion of rest frames\cite{Winicour80}.

We have solved the problem of rest frames in the past with the construction
of the so called nice sections\cite{Moreschi88}. 
Given an asymptotically flat spacetime,
these sections provide a well defined notion of rest frames 
at future null infinity.
We have also proved that they have the expected physical 
properties\cite{Dain00',Moreschi98}(See comments in the next section.).

In this article we use this construction to give a definition of
intrinsic angular momentum in general relativity,
solving in this way the deficiencies of previous 
works\cite{Bramson75,Dray84,Katz97,Moreschi86,Penrose82,Prior77,
Rizzi98,Streubel78,Tamburino66,Wald00,Winicour68}.

%We will make use of the GHP\cite{Geroch73} notation.

\section{Rest frame systems}\label{sec:rest}

In order to define rest frames, we need to fix the notion of supermomentum.
We will use the supermomentum `psi' that was used in the nice section
construction\cite{Moreschi88}.

Let the section $S$ of future null infinity be characterized by the condition
$u=0$, of the Bondi coordinate system $(u,\zeta,\bar\zeta)$. 
Then the supermomentum is given by
\begin{equation} \label{supermomentum}
P_{lm}(S)=-\frac{1}{\sqrt{4\pi}} \int_S Y_{lm}(\zeta, \bar \zeta) 
\Psi(u=0,\zeta,\bar \zeta) dS^2, 
\end{equation}
where $dS^2$ is the surface element of the unit sphere on $S$,
$Y_{lm}$ are the spherical harmonics, the scalar $\Psi$ is given by
\begin{equation} \label{Psi}
\Psi \equiv \Psi_2^0 + \sigma_0 \dot{\bar \sigma}_0 +\eth^2 \bar \sigma_0 ,
\end{equation}
where we are using  the GHP\cite{Geroch73} notation and
where $\Psi_2^0 $ is the leading order asymptotic behavior of the second 
Weyl tensor component,
 $\sigma_0$ is the leading order of  the Bondi shear,  
$\eth$  is the edth operator 
of the unit sphere, and a dot means partial derivative with
respect to the retarded time $u$.

% new paragraph and minor changes
The condition for a section to be of the nice type is that all the
spacelike components of the supermomentum vanish. That is, if $\tilde S$ is
nice then $\tilde P_{lm}(\tilde S)= 0$ for $l \neq 0$.

Since any section $\tilde S$ can be obtained from an arbitrary reference
section $S$ by a supertranslation $\gamma(\zeta,\bar \zeta)$, 
it is useful to know
what is the equivalent condition on $\gamma$. We have shown that
%Let $\gamma(\zeta,\bar \zeta)$ be a supertranslation. 
%Given the section $S$, any
%nice sections can be obtained from $S$ by a
%supertranslation $\gamma$. The
the
condition for the section determined by $\gamma$ to be of the nice
type is\cite{Moreschi88,Moreschi98} 
\begin{equation} \label{nice}
\eth^2 \bar \eth^2  \gamma =\Psi(\gamma,\zeta,\bar \zeta) 
 +K^3(\gamma,\zeta, \bar \zeta) M(\gamma),
\end{equation}
where the conformal factor   $K$ can be related to the Bondi momentum by 
$
K=\frac{M}{P^{\tt a} l_{\tt a}},
$
with
\begin{equation} \label{la}
 (l^{\tt a}) =\left( 1,\frac{\zeta +\bar \zeta }{
1+\zeta \bar{\zeta }},\frac{\zeta -\bar{\zeta }}{i(1+\zeta 
\bar{\zeta )}},\frac{\zeta \bar{\zeta }-1}{1+\zeta \bar{\zeta 
}}\right)
\end{equation}
and $P^{\tt a}$ is evaluated at  the section $u=\gamma$.
Also, the rest mass $M$ at the same section  is given by
$
M=\sqrt{P^{\tt a} P_{\tt a}}
$\cite{Moreschi88,Moreschi98}.

It has been proved that there exists a four parameter family
of solutions of equation (\ref{nice}) with the expected physical 
properties\cite{Dain00',Moreschi98}. 
% new sentence
An important property is that; if $S_1$ is a nice section and
$S_2$ is another nice section which is generated from $S_1$ 
by a timelike translation, then $S_2$ is to the future of $S_1$.
This is exactly what happens in Minkowski space when $S_1$ is
the intersection of the future null cone emanating from an interior point, let
us say $x_1$, with future null infinity; and if $S_2$ is the
intersection of the future null cone corresponding to another point $x_2$
which is in the future of $x_1$. There is then an analogy between rest frames
at future null infinity centered on a nice section $S$ and rest frames of 
Minkowski spacetime centered on a point $x$.

\section{Definition of intrinsic angular momentum}\label{sec:intrinsic}

We will define intrinsic angular momentum by using 
the so called 
`Charge integrals of the Riemann tensor'\cite{Moreschi86}.

A charge integral of the Riemann tensor is a quantity ascribed to a
2-sphere $S$ by the integration of the 2-form  $C_{ab}$, namely:
\begin{equation}
  \label{eq:chargeinte}
  Q_{S} =\int_{S}C 
\end{equation}
where $C_{ab}$ is expressed in terms of the curvature by
\begin{equation}
  \label{eq:chargeint}
  C_{ab} \equiv R_{ab}^{*\;\;cd}\; w_{cd} ,
\end{equation}
and where a right star means right dual of the Riemann tensor and
the 2-form $w_{ab}$ is defined next.
In our case $S$ is assumed to be a nice section\cite{Moreschi88} 
of future null infinity.

%new sentence
How is one supposed to chose the 2-form $w$? It was discussed in the 
literature\cite{Penrose82,Moreschi86} that, in terms of the spinor notation,
it is convenient to require
%\begin{equation}
%  \label{eq:divw}
$
  -\nabla _{A} ^{\;\;B' }\; w^{AB} + {\tt c.c.}= v^{BB'}
$
%\end{equation}
and
%\begin{equation}
%  \label{eq:symw}
$
  \nabla _{E' (E} \; w_{FG)} =0;
$
%\end{equation}
where the vector $v^{BB'} $ is a generator of asymptotic 
symmetries(for the details behind this and other matters see
the full length paper on this subject\cite{Moreschi02}).
However on a non-stationary spacetime, these equations have
in general no solution at future null infinity. In spite of that
we can require this equations to be satisfied on a nice section.
Let $w_{2}, w_{1}$ and $w_{0}$ be the spinor components of the 2-form $w$.
Then we  require
\begin{align}
 w_{2} &=-\frac{1}{3} \bar\eth \bar a  , \label{eq:wsolut1} \\
 w_{1} &=w_{1}^{00} (\zeta ,\bar{\zeta } )+
            \frac{1}{6}\,u\, \eth \bar\eth \bar a  , \label{eq:wsolut2}\\
 w_{0} &=w_{0}^{00} +u\left( -2 \eth w_{1}^{00} 
            + \frac{2}{3} \sigma_0  \bar\eth \bar a 
          \right) - \frac{1}{6}\, u^2 \, \eth^{2} \bar\eth \bar a  \label{eq:wsolut3}
\end{align}
where $a$ is a spin weight 0 quantity satisfying 
$\dot a=0$ and $\bar\eth^2 \bar a =0$,
%$v_{\bar m}= v_{AB'}\,\hat{\iota}^{A} \hat{o}^{B'}$ and 
$w_{1}^{00} $ 
and $w_{0}^{00}$ are spin weight 0 and 1 functions 
respectively that solve the equations
\begin{equation}
  \label{eq:w100}
\eth ^{2} w_{1}^{00} =\frac{1}{3}\eth \sigma_0 \;\bar{\eth}\bar{a} +
\frac{1}{2} \sigma_0\;\eth \bar{\eth}\bar{a} =
-\eth\sigma_0 \;w_{2} -\frac{3}{2} \sigma_0 \;\eth w_{2} 
\end{equation}
and
\begin{equation}
  \label{eq:w000}
\eth  w_{0}^{00} =-2\sigma_0 w_{1}^{00} .
\end{equation}
Using the potential $\delta$ of the shear satisfying
$
  \sigma_0 = \eth^2 \delta
$,
the component $w_1$ can be expressed by
\begin{equation}
  \label{eq:w1alfa}
  w_1 = b + \frac{1}{3} \eth \delta \bar\eth \bar a 
+ \frac{1}{6} (u - \delta) \eth \bar\eth \bar a ;
\end{equation}
where the spin weight 0 quantity $b$ satisfies $ \dot b =0$ and $\eth^2 b = 0$.

Having a timelike one-parameter family of nice sections at future null infinity,
this procedure provides with
a 2-form $w$ at future null infinity with the functional dependence
\begin{equation}
  \label{eq:w0}
  w_{AB}^{} =w_{AB}^{} \left( \sigma_0(u,\zeta ,\bar{\zeta } ),a,b
;u,\zeta ,\bar{\zeta } \right) .
\end{equation}

Then the charge integral becomes
\begin{equation}
  \label{eq:charge}
\begin{split}
Q_S(w)=  \int_{S}C  \\
 =4\int_{S}\Bigl( -& \, w_{2} \left( \Psi_{1}^{0} 
+ 2\sigma_0 \eth\bar{\sigma_0 } +\eth\left( \sigma_0 \bar{\sigma_0 } \right)
\right)  \Bigr. \\
+ \Bigl.
& \, 2 \,w_{1} \left( \Psi _{2}^{0} +\sigma_0 \dot{\bar{\sigma_0 } } +
\eth^{2} \bar{\sigma_0 } \right) \Bigr) \;dS^{2} + {\tt c.c.} .
\end{split}
\end{equation}

In order to pick up the intrinsic angular momentum we need to
impose the center of mass condition\cite{Moreschi02}
\begin{equation}
  \label{eq:acondit2}
  Q_S(a)=0 \qquad \text{for all}\quad a=\bar a .
\end{equation}
This is the analog of the fixing of the
translation freedom in the Minkowskian case
by requiring that three components of the angular
momentum tensor vanish, namely
$J^{01}=J^{02}=J^{03}=0$.
In this way, there is left
a unique one parameter family of nice sections,
which determines the center of mass frames.

Let us denote these sections with $S_{\tt cm}$;
then the intrinsic angular momentum $j$ is defined by
\begin{equation}
  \label{eq:instrin}
  j(w)=Q_{S_{\tt cm}}(w) ;
\end{equation}
where in order to pick up the intrinsic angular momentum one must
take $a=-\bar a$ and $b=0$\cite{Moreschi02}.

\section{Comments}
\begin{figure}[htbp]
\centering
\includegraphics[clip,width=0.45\textwidth]{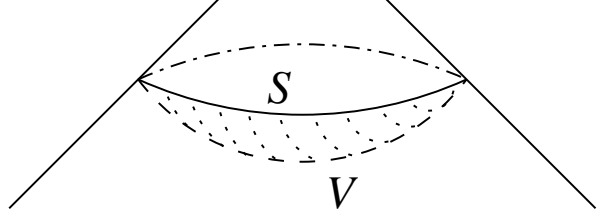}
\caption{The spacelike hypersurface $V$
has as boundary the section {\it S} at future null infinity.
}
\label{fig:inte}
\end{figure}

Several advantages exist when one uses the charge integral approach
for the definition of physical quantities.
For example, although we have defined the tensor $C$ at future null
infinity, let us consider the possibility to extend its definition
to the interior of the spacetime. Assume $V$
is a spacelike hypersurface in the interior of the spacetime which
extends up to future null infinity, and which has as boundary
the section $S$; as shown in figure \ref{fig:inte}.

Then, using Stokes' theorem, one can express the charge integral
$Q$ as an integral on $V$, namely
\begin{equation}
  \label{eq:qonsigma}
    Q_{S} =\int_{S}C = \int_{V} dC .
\end{equation}
One can prove\cite{Moreschi02} that
the exterior derivative of $C$ can be expressed by
\begin{equation}\label{eq:dC2}
\begin{split}
dC_{abc}  = & \frac{1}{3} \epsilon _{abcd} \ ^{*}R^{*defg} \ \left(
T_{efg} +\frac{1}{3} g_{ef} v_{g} -\frac{1}{3} g_{eg} v_{f} \right)  \\
= & \frac{1}{3} \epsilon _{abcd} \left( -2G^{dg} v_{g} + {}^{*}R^{*defg}
\ T_{efg} \right) ,
\end{split}
\end{equation}
where $G_{ab}$ is the Einstein tensor,
$T_{abc}$ is the traceless part of $\nabla_a w_{bc}$ and
$\nabla_a w^{ab}=v^b$, its trace; in other words
\begin{equation}
  \label{eq:nablaw}
  \nabla _{a} w_{bc} = T_{abc} +\frac{1}{3} g_{ab} v_{c} -\frac{1}{3}
g_{ac} v_{b} .
\end{equation}

Let us look at equation (\ref{eq:dC2}) in the linearized gravity case.
Then, it is noted\cite{Penrose82} that  if the vector $v^a$
were a Killing vector of the flat background metric, and $T_{abc}$
were $O(1)$, then equation (\ref{eq:qonsigma}) will give the
conserved quantities in the context of linearized gravity. This
is telling us that the charge integrals have the appropriate
meaning in the weak field limit.

\begin{figure}[htbp]%\label{fig:flux}
\centering
\includegraphics[clip,width=0.45\textwidth]{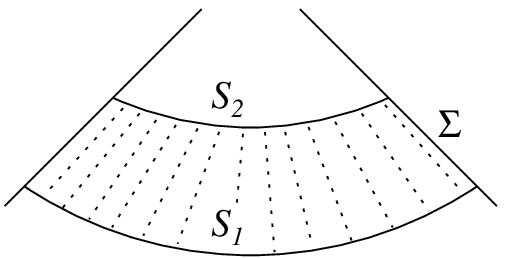}
\caption{The null  hypersurface $\Sigma$ at future null infinity
has as boundary the sections $S_1$ and $S_2$.
}
\label{fig:flux}
\end{figure}

Also one can use Stokes' theorem to calculate the flux of angular momentum.
Let $S_2$ be a section to the future of the section $S_1$ of future null infinity, 
and let
now $\Sigma$ be the region which has as boundaries  $S_1$ and $S_2$;
see figure \ref{fig:flux}. Then the
flux law for angular momentum is given by
\begin{equation}
  \label{eq:fluxq}
    Q_{S_2}-Q_{S_1}  =\int_{S_2}C - \int_{S_1}C = \int_{\Sigma} dC .
\end{equation}

When the spacetime is stationary, it can be seen\cite{Moreschi02} that
our definition does give the intrinsic angular momentum of the spacetime;
and also that the center of mass, defined in terms of nice sections, is what
one expects to be.

Let us consider a radiative spacetime in which
one can distinguish three stages; starting with a stationary
regime, passing through a radiating stage and ending in a stationary regime.
One can see that our construction gives the correct
intrinsic angular momentum in the first and third stages; 
even though the center of mass frames
of the first and third stages are related in general by a supertranslation.

It should be stressed that given a point at future null infinity, which it
could be considered as the event of the detection of gravitational waves,
then the center of mass sections just defined 
%together with the intrinsic angular momentum, 
provides with a unique section where all physical quantities should
be defined. A couple of previous works\cite{Moreschi86,Rizzi98} 
also provide with a prescription
for the selection of unique sections at future null infinity,
however they are either non-local\cite{Moreschi86}
or they do not mention the subject of supermomentum\cite{Rizzi98}
and therefore can not be considered as either rest frames nor center of mass.

It is probably worth mentioning that several estimates on the gravitational 
radiation of different astrophysical systems are done with the quadrupole
radiation formula. If one needs these estimates at two different 
times, such that back reaction has occurred due to the emission of gravitational
waves between the first and second time, then the multipoles should be 
calculated at the corresponding center of mass for each time. 
In other words, even in the use of the quadrupole radiation formula one
needs to have a well defined notion of center of mass. Our work provides
both notions `center of mass' and `intrinsic angular momentum' 
in one construction.

\section*{Acknowledgments}
We would like to thank Dr. A. Rizzi for ongoing discussions on his work
and Dr. C. Kozameh for help in these matters.
We acknowledge support from SeCyT-UNC and CONICET.

\end{document}